\documentclass[twocolumn,rmp,showpacs]{revtex4}

\usepackage{dcolumn,graphicx,amsmath,amssymb}
\usepackage{txfonts}

\begin{document}

\title{Dynamics of networking agents competing for high centrality and
  low degree}
\author{Petter Holme}
\author{Gourab Ghoshal}
\affiliation{Department of Physics, University of Michigan, Ann Arbor,
MI 48109, U.S.A.}

\begin{abstract}
We model a system of networking agents that seek to optimize their
centrality in the network while keeping their cost, the number of
connections they are participating in, low. Unlike other game-theory
based models for network evolution, the success of the agents is
related only to their position in the network. The agents use
strategies based on local information to improve their chance of
success. Both the evolution of strategies and network structure are
investigated. We find a dramatic time evolution with cascades of
strategy change accompanied by a change in network structure. On
average the network self-organizes to a state close to the transition
between a fragmented state and a state with a giant
component. Furthermore, with increasing system size both the average
degree and the level of fragmentation decreases. We also observe that
the network keeps on actively evolving, although it does not have to,
thus suggesting a Red Queen-like situation where agents have to keep
on networking and responding to the moves of the others in order to
stay successful.
\end{abstract}


\maketitle

\section{Introduction}

Game theory conceptualizes many of the circumstances that drive the
dynamics of social and economic systems. If such systems consist of
many pair-wise interacting agents they can be modeled as networks. In
such networks one can relate the function of a vertex to its position.
For example, in business connections an agent would presumably like to
be close, in network distance, to the average other
agent~\cite{harary,wf}. This ensures the information received from
other agents to be up to date~\cite{rosvall1} and will likely increase
the agent's sphere of influence. At the same time the agent would 
seek to limit the work load by minimizing its degree (number of
connections). In this paper we define an iterative $N$-player game
where agents try simultaneously to obtain high centrality and low
degree. Agents remove and add edges by individual
strategies. Furthermore they update the strategies throughout the game
by imitating successful agents. We assume the agents have only
information about their immediate surroundings. As a result an agent
can only re-link to, or observe and mimic the strategies of, other
agents a fixed distance away.
Most recent studies of games on networks, have considered a static
underlying network defining the possible competitive
encounters~\cite{vukov:pd,our:realpd,our:pd,wu:pd,santos:pd}.
In other models where the network co-evolves with the
game~\cite{egui:pre,egui:ajs,ebelbornholdt:coevo}, 
the agents are assigned additional variables which serve as the basis
of the game. In our model however, the score of an agent is determined
by the network dynamics alone. This setting, apart from being
conceptually simpler, makes the relation between the game and network
dynamics more transparent.
The rest of the paper contains a precise definition of the model, an
investigation of the time evolution of the strategies and network
structure, and an investigation of the dependence on model
parameters.

\hfill

\section{Definition of the game}

\subsection{Score and moves}

In our model $N$ agents are synchronously updated over
$t_\mathrm{tot}$ iterations. The initial configuration is an
Erd\H{o}s-R\'{e}nyi network~\cite{er:on} of $M_0$ edges. All steps of
the dynamics keep the network simple so that as multiple edges or
self-edges do not occur.
The score, in our game, is an effective score taking into account both
the benefit of centrality and the inevitable cost of maintaining the
network ties. We want the score of a vertex $i$ to increase with
centrality and decrease with its degree $k_i$. Of many centrality
concepts~\cite{harary,wf} we choose to base our score on the simplest
non-local centrality measure---closeness centrality (the reciprocal
average path-length from one vertex to the rest of the
graph). Furthermore, if the network is disconnected we would like the
score to increase with the number of vertices reachable from $i$. To
incorporate this we use a slight modification of closeness
\begin{equation}\label{eq:cent}
  c(i)=\sum_{j\in H(i)\setminus\{i\}} \frac{1}{d(i,j)},
\end{equation}
where $H(i)$ is the connected subgraph $i$ belongs to and $d(i,j)$ is
the graph distance between $i$ and $j$. The number of elements in the
sum of Eq.~(\ref{eq:cent}) is proportional to the number of vertices
of $i$'s connected component. We use the average reciprocal distance,
rather than the reciprocal average distance. The former gives a
higher weight on the count of closer vertices, but captures similar
features as the original closeness does. We define a score function
that incorporates the desired properties mentioned above:
\begin{equation}\label{eq:score}
  s(i) = \left\{\begin{array}{ll}c(i)/k_i & \mbox{if
        $k_i>0$}\\ 0 & \mbox{if $k_i=0$}\end{array}\right. .
\end{equation}
In addition we assume the accessible information is restricted to a
close neighborhood of a vertex. To be precise, the moves allowed to a
vertex is to delete or add edges to agents up to two steps
away. Our assumption is motivated by the fact that in real world
systems, agents are more likely to have knowledge of a restricted
fraction than the whole network itself.

\subsection{Strategies}

When a vertex $i$ updates its position, it selects another vertex
in a set $X$ (the neighborhood $\Gamma(i)$ if an edge is to be
removed, or the second neighborhood $\Gamma_2(i)=\{j:d(i,j)=2\}$ if an
edge is to be added). This is done by successively applying six
tiebreaking \textit{actions}:
\begin{itemize}
\item Choose vertices with maximal (minimal) degree (MAXD / MIND).
\item Choose vertices with maximal (minimal) centrality in the
  sense of Eq.~(\ref{eq:cent}) (MAXC / MINC).
\item Pick a vertex at random (RND).
\item Do not add (or remove) any edge (NO).
\end{itemize}
The strategies of a vertex is encoded in two six-tuples
$\mathbf{s}_\mathrm{add}=(s^\mathrm{add}_1,\cdots,s^\mathrm{add}_6)$
and $\mathbf{s}_\mathrm{del}$ representing a priority ordering of the
addition and deletion actions respectively. If $\mathbf{s}_\mathrm{add}(i)=
(\mathrm{MAXD},\mathrm{MINC}, \mathrm{NO},\mathrm{RND},
\mathrm{MIND},\mathrm{MAXC})$ then $i$ tries at first to attach an
edge to the vertex in $\Gamma_2(i)$ with highest degree. If more than
one vertex has the highest degree, then one of these is selected by
the MINC strategy. If still no unique vertex is found, nothing is
done (by application of the NO strategy). Note that such a vertex
is always found after strategies NO or RND are applied. If
$X=\varnothing$ no edge is added (or deleted).

\begin{figure}
  \resizebox*{\linewidth}{!}{\includegraphics{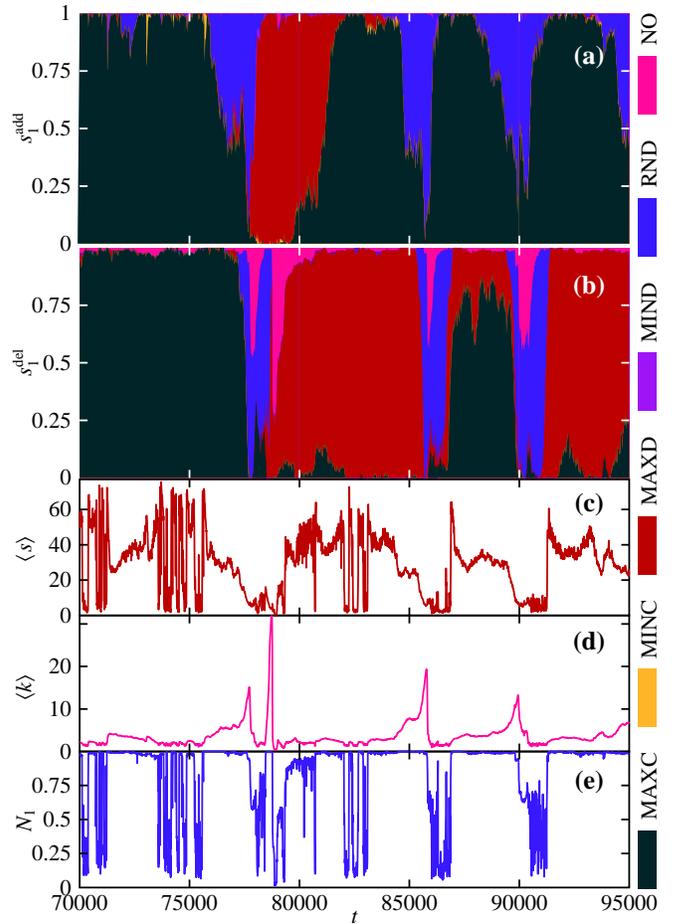}}
  \caption{ An example run for a 200-agent system with
    $p_r=0.012$. (a) and (b) show the fraction of vertices having a
  certain leading strategy for addition (a) and deletion (b)
  respectively. (c) show the average score, (d) the average degree and
  (e) the number of vertices in the system.
  }
  \label{fig:evo}
\end{figure}

\subsection{Strategy updates and relinking errors}

The strategy vectors are initialized to random permutations of the six
actions. Every $t_\mathrm{strat}$'th time step a vertex $i$ updates
its strategy vectors by identifying the vertex in
$\Gamma_i\cup\{i\}=\{j:d(i,j)\leq 1\}$ with highest accumulated score
since the last strategy update. Then $i$ copies the parts of
$\mathbf{s}_\mathrm{add}(j)$ and $\mathbf{s}_\mathrm{del}(j)$ that $j$
used the last time step, and let the remaining actions come in the
same order as the strategy vectors prior to the update.
For the purposes of making the set of strategy vectors ergodic,
drive the strategy optimization~\cite{nowak:wsls,lindgrennordahl} and
model irrational moves by the
agents~\cite{kah:bounded_rationality} we swap, with probability $p_s$,
two random elements of $\mathbf{s}_\mathrm{add}(j)$ and
$\mathbf{s}_\mathrm{del}(j)$ every strategy vector update.
Like the strategy space we also want the network space to be ergodic
(i.e.\ that the game can generate all $N$-vertex graphs from all initial
configurations). In order to ensure ergodicity disconnected clusters should
be able to be re-connected. We obtain this by letting a vertex $i$
attach to a random vertex (not just a second neighbor) with
probability $p_r$ every $t_\mathrm{rnd}$'th time step. This is also
plausible in real socioeconomic networks---even if agents are more
influenced by their network surrounding, long-range links can form by
other mechanisms (cf.\ Ref.~\cite{wattsstrogatz}).

\subsection{The entire algorithm}

The outline of the algorithm is thus:
\begin{enumerate}
\item\label{step:init_nwk} Initialize the network to an
  Erd\H{o}s-R\'{e}nyi network with $N$ vertices and $M$ edges.
\item\label{step:init_s} Use random permutations of the six actions
  as $\mathbf{s}_\mathrm{add}$ and $\mathbf{s}_\mathrm{del}$ for all
  vertices.
\item\label{step:score} Calculate the score for all vertices.
\item\label{step:rewi} Update the vertices synchronously by adding
  and deleting edges as selected by the strategy vectors. With
  probability $p_r$ an edge is added to a random vertex instead of a
  neighbor's neighbor.
\item\label{step:strat} Every $t_\mathrm{strat}$'th time step, update
  the strategy vectors. For each vertex, with probability $p_s$, swap
  two elements in a vertex' strategy vector.
\item\label{step:iter} Increment the simulation time $t$ and, if
  $t<t_\mathrm{tot}$, go to step~\ref{step:score}.
\end{enumerate}
$n_\mathrm{avg}$ averages over different realizations of the algorithm
are performed. We will use parameter values $M_0=3N/2$, $p_s=0.005$,
$t_\mathrm{strat}=10$, $t_\mathrm{tot}=10^5$ and $n_\mathrm{avg}=100$
throughout the paper (the conclusions will not depend sensitively on
these values).

\section{Time evolution}

A part of the time evolution of a run of the game is displayed in
Fig.~\ref{fig:evo}. Fig.~\ref{fig:evo}(a) and (b) show the fraction
of the agents having a specific main addition ($s^\mathrm{add}_1$) and
deletion action ($s^\mathrm{del}_1$) respectively. As we can see, the
time evolution can be very complex, having sudden cascades of
strategy changes. We do not display actions with lower priorities
($s_2,\cdots,s_6$), but we note that they
are less clear-cut as they experience a lower selection pressure. 
Typically the time evolution shows rather lengthy
quasi-stable periods punctuated by outbursts of strategy changing 
cascades (in both the addition and deletion strategies) as seen in
Fig.~\ref{fig:evo}(a) and (b). Not all strategies, as we will see
later, invade the population. As illustrated in this example, MAXC is the most
frequent main action for most parameter values, whereas MINC and
MIND (and NO for addition) are rare. From the definition of the
actions we anticipate differences in the network structure for time
frames of different dominating strategies. This is indeed the case as
evident from panels (c), (d) and (e) of Fig.~\ref{fig:evo} which display
the average score $\langle s\rangle$, degree $\langle
k\rangle$ and number of vertices in the largest connected cluster
$n_1$. The average score fluctuates wildly suggesting that states of
global prosperity are unstable. Likewise the degree has an
intermittent time evolution with sudden high-degree spikes and periods
of sparseness. Unsurprisingly, the high-degree spikes are located at
the outbursts of the NO deletion strategy where edges are not deleted,
but only added. The size of the largest connected cluster has an even more
dramatic time evolution, fluctuating between fully connected and
fragmented states. Note that there need not be a dramatic change in
degree to initiate a drop in $n_1$---this leads us to conclude that the 
phenomenon probably arises from network topological effects.

\begin{table*}
\begin{ruledtabular}
\begin{tabular}{r|cccccc|cccccc}
   & \multicolumn{6}{c}{addition} &
  \multicolumn{6}{c}{deletion} \\
& MAXC & MINC & MAXD & MIND & RND & NO & MAXC & MINC & MAXD & MIND &
RND & NO\\\hline
MAXC & 1 & 0.0164(3) & 0.0088(2) & 0.0107(4) & 0.0151(5) & 0.0010(0) &
1 & 0.0100(2) & 0.0131(4) & 0.0094(2) & 0.0266(3) & 0.0126(3)\\
MINC & 0.0169(3) & 1 & 0.0113(6) & 0.036(2) & 0.025(2) & 0.0017(3) &
0.0098(2) & 1 & 0.0070(3) & 0.010(1) & 0.0105(4) & 0.0050(3) \\
MAXD & 0.0093(3) & 0.0104(7) & 1 & 0.0103(6) & 0.0206(9) & 0.0003(0) &
0.0133(4) & 0.0067(3) & 1 & 0.0055(2) & 0.0124(3) & 0.0062(2)\\
MIND & 0.0115(4) & 0.030(2) & 0.0130(7) & 1 & 0.059(5) & 0.0020(2) &
0.0087(2) & 0.011(1) & 0.0054(2) & 1 & 0.0101(2) & 0.0055(3)\\
RND & 0.0157(5) & 0.024(2) & 0.020(1) & 0.064(5) & 1 & 0.0023(5) &
0.0269(3) & 0.0094(4) & 0.0128(3) & 0.0083(2) & 1 & 0.0072(3)\\
NO & 0.0007(0) & 0.0031(2) & 0.0009(0) & 0.0036(2) & 0.0042(4) & 1 &
0.0097(3) & 0.0076(3) & 0.0053(2) & 0.0078(3) & 0.0131(3) & 1\\
\end{tabular}
\end{ruledtabular}
\caption{Values for the $\mathbf{\Theta}$ matrices for addition and
  deletion. ($\Theta_{ij}$ is the deviation from the expected value in
  a model of random transitions given the diagonal values.)
  The values are averaged over 100 realizations of the
  algorithm. All digits are significant to one s.d. The parameter
  values are the same as in Fig.~\ref{fig:evo}. Numbers in parentheses
  are the standard errors in units of the last decimal.}
\label{tab:trans}
\end{table*}

Note that in Fig.~\ref{fig:evo}(b) the
strategies seem to differ in their ability to invade one another,
e.g.\ MAXC is followed by a peak in RND. We investigate this
qualitatively by calculating the ``transition matrix'' $\mathbf{T}$
with elements $T(s_1,s_1')$ giving the probability of a vertex with
the leading action $s_1$ to have the leading action $s_1'$ at the
next time step. However note that the dynamics is not fully determined
by $\mathbf{T}$, and is thus not a transition matrix in the sense of
other physical models. If that were the case (i.e.\ the current strategy is
independent of the strategy adopted in the previous time step) we would have
the relation $T_{ij}=\sqrt{T_iT_j}$. So we measure the deviation from
such a null-model by assuming the diagonal (i.e.\ the frequencies of
the strategies) and calculating
$\mathbf{\Theta}$ defined by
\begin{equation}\label{eq:theta}
  \Theta_{ij} = T_{ij} / \sqrt{T_iT_j}.
\end{equation}
The values of $\mathbf{\Theta}$ for the parameters defined in
Fig.~\ref{fig:evo} are displayed in Tab.~\ref{tab:trans}. The
off-diagonal elements are much lower than $1$ (the average off-diagonal
$\Theta$ values are $0.014$ for addition strategies and $0.010$ for
deletion). This reflects the contiguous periods of one dominating
action. Note that transitions between MAXC and RND are
over-represented: $\Theta^\mathrm{del}_{\mathrm{MAXC},
  \mathrm{RND}}\approx \Theta^\mathrm{del}_{\mathrm{RND},
  \mathrm{MAXC}}\approx 0.027$, which is more than twice the value of
any other off-diagonal element involving MAXC or RND. To add to the
complexity, the matrix is not completely symmetric
$\Theta^\mathrm{del}_{\mathrm{RND}, \mathrm{NO}}$ is twice ($\sim 3$
s.d.)\ as large as $\Theta^\mathrm{del}_{\mathrm{NO}, \mathrm{RND}}$
meaning that it is easier for RND to invade NO as a leading deletion
action, than vice versa.

\begin{figure}
  \resizebox*{\linewidth}{!}{\includegraphics{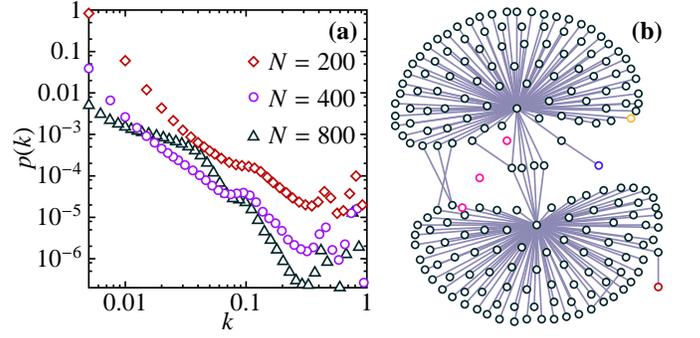}}
  \caption{ The structure of the network with MAXC as the leading
    addition action. Parameter values are the same as in
    Fig.~\ref{fig:evo}. Part (a) shows the degree distribution for time steps
    where more than a half of the agents use MAXC as leading
    action. The curves are log-binned for large degrees. Errors are
    smaller than the symbol size. Part (b) shows an example graph for
    $N=200$. We emphasize that this is only one
    of a great variety on network topologies that emerge from the
    dynamics. The colors of the vertices represent the addition actions
    as in Fig.~\ref{fig:evo}.
  }
  \label{fig:d}
\end{figure}

\section{Degree distributions and the influence of degree on score}

To get a more detailed view of the relation between the preferred
actions and the structure of the network, we investigate the degree
distribution $p(k)$ for different leading actions. In
Fig.~\ref{fig:d}(a) we plot the degree distribution for the MAXC
dominating addition action. It is conspicuously wide---so despite the
fact that the vertex strategies are similar, the network structure
evolves into a highly inhomogeneous state. There are peaks in the
degree distribution close to $k\approx 0.4 N$ and $k\approx 0.8 N$,
meaning that the network has at least one or more hubs of extremely
high degree. A snapshot of the network with two hubs, each with degree
close to $N/2$ is seen in Fig.~\ref{fig:d}(b). Such a situation can
indeed be rather stable: The most central vertices (the vertices
between the hubs) have rather low degree, and thus have a very high
score. Since these are in $\Gamma_2$ (but not in $\Gamma$) of most
vertices, these will be the hubs of the next time step, and the old
hub will likely be between these. Thus the property of being a hub
will effectively oscillate between members of two sets of vertices.

\begin{figure}[b]
  \resizebox*{\linewidth}{!}{\includegraphics{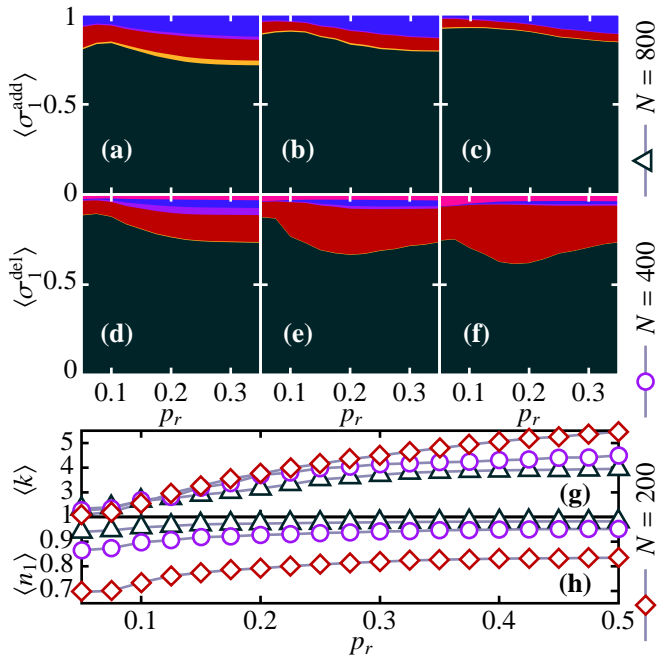}}
  \caption{ The system's dependence on the fraction of random
    rewirings $p_r$ and system size $N$. Parts (a), (b) and (c) show the
    fraction of preferred addition actions $\langle
    \sigma_1^\mathrm{add} \rangle$ for systems of 200, 400 and 800
    agents respectively. Parts (d), (e) and (d) show the fraction of preferred
    deletion actions for the same three system sizes, while (g) shows the
    average degree and (h) the average size of the largest connected
    component.}
  \label{fig:r}
\end{figure}

\section{Dependence on system size and error rates}

Next we turn to the scaling of the strategy preferences and structural
measures with respect to model parameters. In Fig.~\ref{fig:r} we tune
the fraction of random attachments $p_r$ for three system sizes. In
panels (a)-(c) we display the fraction of leading addition actions
among the agents $\langle \sigma_1^\mathrm{add} \rangle$ (averaged over $\sim
100$ runs and $10^5$ time steps). As observed in Fig.~\ref{fig:evo}(a)
the dominant strategy is MAXC followed by MAXD and RND. The leading
deleting actions, as seen in panels (d)-(f), are ranked similarly
expect that MAXD has a larger (and increasing) presence. There are
trends in the $p_r$-dependences of $\langle \sigma_1^\mathrm{add}
\rangle$, but apparently no incipient discontinuity. This observation
(which also seems to hold for $p_s$ scaling) is an indication that the
results above can be generalized to a large parameter range. We also
note that although the system has the opportunity to be passive (i.e.\
agents having $s_1^\mathrm{add} = s_1^\mathrm{del} =\mbox{NO}$), it
does not. This is reminiscent of the ``Red Queen hypothesis'' of
evolution~\cite{redqueen}---organisms need to keep evolving to
maintain their fitness. The average degree, plotted in
Fig.~\ref{fig:evo}(g) is monotonously increasing with $p_r$ and
decreasing with $N$ (if $p_r\gtrsim 0.12$). For all network models
that we are aware of (allowing for fragmented networks) decreasing
average degree implies a smaller giant component. In our model the
picture is the opposite, as the system grows the giant component spans
an increasing fraction of the network. This also means that the agents
collectively reach the twin goals of keeping the degree low and the
graph connected.

\section{Summary and conclusions}

To summarize, we have investigated an $N$-player game of networking
agents. The success of an agent $i$ increases with the closeness
centrality and the size of the connected component $i$ belongs to, while it
decreases with $i$'s degree. Such a situation may occur in diplomacy,
lobbying or business networks, where an agent wants to be central in
the network (for the purpose of having as new information as possible
and be more actively involved in the decision making process) but not
at the expense of having too many direct contacts. The dynamics
proceed by the agents deleting edges and attaching new edges to their
second-neighbors according to strategies based on local
information. Once in a while (every tenth time step in our simulation)
the agents evaluate the strategies of the neighborhood and imitate the
best performing neighbor to optimize their strategy. As the vertices
of our model have no additional traits---their competitive situation
is completely determined by their network position---the time
evolution of strategies is immediately tied to the evolution of network
structure. These evolutionary trajectories are strikingly complex
having long periods of relative stability followed by sudden
transitions, spikes, or chaotic periods visible in both the 
strategies and the network structure. One such instability is
manifested in a transient fragmentation of the network, this occurs
more rarely as the network size increases. In fact the network gets more
connected as size is increased, interestingly this is accompanied with
a decreasing fraction of links---thus, with a growing number of actors
the system gets better at achieving the common goal of being connected
and keeping the degree low. We also observe that the network dynamics
never reaches a fixed point of passivity (where the network is largely
static), this suggests situation similar to the Red Queen
hypothesis---agents have to keep on networking to maintain their
success. We believe network positional games will prove to be a useful
framework for modeling dynamical networks, and anticipate much future
work in this direction.

\begin{acknowledgements}
  The authors thank Mark Newman for comments. P.H. acknowledges
  financial support from the Wenner-Gren foundations.
\end{acknowledgements}

\end{document}